\documentstyle[preprint,aps,prb]{revtex}
\newcommand{\ds}{\displaystyle}

\newcommand{\non}{\nonumber}

\newcommand{\ii}{{\rm i}}
\newcommand{\kpara}{{k_{y}}}

\newcommand{\e}{\varepsilon}

\newcommand{\k}{{\mbox{\boldmath$k$}}}

\newcommand{\be}{\begin{eqnarray}}
\newcommand{\en}{\end{eqnarray}}
\begin{document}
\draft
\preprint{condmat/}
\title{Anisotropic suppression of quasiparticle weight  in   two-dimensional electron  system with 
partially flat  Fermi surface: two-loop renormalization-group analysis
}
\author{Jun-ichiro Kishine\thanks{E-mail:kishine@ims.ac.jp} and Kenji Yonemitsu}
\address{Institute for Molecular Science,
Okazaki 444-8585, Japan
}
\date{\today}
\maketitle
\begin{abstract}
Two-loop renormalization-group analysis for
a two-dimensional electron system with a partially flat Fermi surface has been carried out.
We found that, {\it irrespective of pairing mechanism},  the  quasiparticle weight is anisotropically suppressed    due  to  logarithmically singular processes  
in the flat regions of the Fermi surface. 
When the energy scale decreases,  the quasiparticle weight is the most strongly suppressed  around the center of the flat region, which
 qualitatively agrees with the
anisotropic pseudogap behavior suggested through the angle-resolved photoemission
     spectroscopy  experiments for underdoped Bi$_2$Sr$_2$CaCu$_2$O$_8$.

\end{abstract}
%\pacs{Valid PACS appear here.
%{\tt$\backslash$\string pacs\{\}} should always be input,
%even if empty.}
 
\pagebreak
\baselineskip18pt

Non-Fermi-liquid behavior in the normal state of high $T_c$ superconductors has provoked a great deal of
controversy. \cite{Imada98}  Especially, the pseudogap phenomena in the underdoped region has been one of the  main points at issue.
Recent angle-resolved photoemission spectroscopy (ARPES) experiments on  
 underdoped Bi$_2$Sr$_2$CaCu$_2$O$_8$\cite{Norman98,Ding96} have revealed evidence  that on cooling
the pseudogap determined through the leading edge of   the ARPES spectra 
first opens up   around the points $(\pm\pi,0)$ and $(0,\pm\pi)$ in the Brillouin zone below some crossover temperature 
$T^\ast\sim180$K without any broken symmetry.  Then,  the Fermi surface 
is broken up into the disconnected 
arcs which  shrink with decreasing temperature above the superconducting transition temperature $T_c\sim 85$K.   

Self-consistent treatments of  the $d$-wave pairing fluctuation  above $T_c$
showed
that the  Fermi surface is broken up into the patches   around  $(\pm\pi,0)$ and $(0,\pm\pi)$.\cite{Engelbrecht97}
Resonating valence bond picture also predicts  the patchy Fermi surface   due to
spinon pairing.\cite{Wen96}  Here we propose another mechanism. 

One aspect to study   interacting fermion systems is to see how  physical quantities
{\it flow} toward their low-energy asymptotic behavior by means of
  the   renormalization-group (RG)  framework.\cite{Shankar94}  
Since the ARPES experiments  detect the coherent one-particle  excitations, 
the key issue here    is   the  quasiparticle weight, $z(\k)$, of an electron with a momentum $\k$.  
To see the RG flow of   the quasiparticle weight, we have to carry out the RG analysis at least up to   a {\lq\lq}two-loop{\rq\rq} order,
since the lowest order perturbative correction to the quasiparticle weight comes from the 
logarithmically singular two-loop self energy diagram.\cite{Solyom79}
In this work,  by applying  the  two-loop RG  analysis to a two-dimensional (2D) interacting electron system with a
partially flat Fermi surface,  we present a new scenario, {\it irrespective of pairing mechanism}, that 
the  quasiparticle weight is anisotropically suppressed  due to  logarithmically singular processes  
in the flat regions of the Fermi surface. 

We consider a 2D electron system with the model Fermi surface
  shown in Fig.~1 
which resembles the real Fermi surface of Bi$_2$Sr$_2$CaCu$_2$O$_8$\cite{Ding96,Marshall96} and
consists of   flat regions ($\alpha$, $\bar \alpha$,
$\beta$,   $\bar \beta$) with  length $2\Lambda$ and  round-arc regions ($A$, $\bar A$, $B$, $\bar B$).  
The centers of the flat regions correspond to $(\pm\pi,0)$ and $(0,\pm\pi)$ points in Bi$_2$Sr$_2$CaCu$_2$O$_8$.
The same model Fermi surface has been studied within the one-loop RG  analysis (parquet approach)
by Zheleznyak, Yakovenko and Dzyaloshinskii.\cite{Zheleznyak97}

 Following the standard RG technique,\cite{Shankar94} we consider  the linearlized  one-particle dispersion around 
the flat regions of the  Fermi surface with bandwidth cutoff $E_0$.
We assume $E_0\ll v_F\Lambda$ with $v_F$ being the correspondign Fermi velocity.
As in the case of one-dimensional (1D) metals,\cite{Solyom79}
 the electron one-particle and two-particle processes within    the parallel flat regions
give rise to the logarithmically singular contribution,
$\ln[E_0/\omega]$ with $\omega\,(\leq E_0)$ being  an infrared cutoff energy scale,
to  the two-particle scattering strengths and the
quasiparticle weight.
On the other hand,   the processes involving at least one electron in the 
round-arc regions and the processes involving two electrons in two flat regions perpendicular to each other, for example $\alpha$  and  $\beta$,
never give rise to this type of singular contribution.
Consequently, for an electron located in the round-arc regions,
the quasiparticle weight $z(\k)$ remains marginal.
Thus only the processes which are within  the parallel flat regions are relevant to
the RG flow of $z(\k)$.\cite{Zheleznyak97}

We start with the action for the electron one-particle and two-particle processes in  the  parallel flat regions, $\alpha$ and $\bar \alpha$,
$S_l=S_{{\rm kin};l}+S_{{\rm int};l}$, where $l$ is a scaling parameter defined below.
The kinetic action  is given by
\begin{eqnarray}
S_{{\rm kin};l}=\sum_{\nu=\alpha,\bar\alpha}\sum_{\sigma}
\int_{-\infty}^{\infty}{d\e\over 2\pi}
\ds\int_{{\cal C}_{\nu;l}}{d\k\over (2\pi)^2}
{\cal G}_{\nu}^{-1}(k_{x},\ii\e)  c_{\nu\sigma}^{\ast}(K)  c_{\nu\sigma}(K),
\end{eqnarray}
where $c^\ast_{\nu\sigma}$ and $c_{\nu\sigma}$ are  the Grassmann variables 
representing  the   electron with the spin $\sigma$ in the flat region $\nu=\alpha\,\,{\rm or}\,\,\bar \alpha$ and
$K=(\k,\e)$ with  $\e$ being a Fermion thermal frequency.
The non-interacting one-particle propagator is given by
$
{\cal G}_\nu^{-1}(k_x,\ii\e)=\ii\e-\varepsilon_\nu(k_x)
$
where the linearized one-particle dispersions are given by
$\varepsilon_\alpha(k_x)=v_F(k_x-k_F)$  and $\varepsilon_{\bar\alpha}(k_x)=v_F(-k_x-k_F)$ 
with $v_F$ and $k_F$ denoting the corresponding Fermi velocity and 
the Fermi momentum, respectively.  We neglect the dependence of the Fermi velocity on $k_y$ which is irrelevant
in the RG sense.
We parametrize the infrared cutoff energy as $\omega_l=E_{0}e^{-l}$ with a scaling parameter, $l$.
Two-dimensional electron momenta, $\k$, are restricted to the set  
${\cal C}_{\nu;l}\equiv\left\{\k \mid \mid \varepsilon_\nu(k_x)
\mid\leq \omega_{l}/2, -\Lambda\leq k_y\leq \Lambda \right\}$.

The action for the   two-particle  processes are written as
\widetext
\begin{eqnarray}
S_{{\rm int};l}&=&{\pi v_{F}}\sum_{\sigma_1,\dots,\sigma_4}
\prod_{i=1,2,3}
\int_{-\infty}^{\infty}{d\e_i\over 2\pi}
\int_{{\cal C}_{l}}{d\k_{i}\over (2\pi)^2}\\
 && g_l^{\sigma_1\sigma_2\sigma_3\sigma_4}(\kpara_1,\kpara_2,\kpara_3)
c_{\bar\alpha\sigma_4}^{\ast}(K_4)  
c_{\alpha\sigma_3}^{\ast}(K_3)  
c_{\alpha\sigma_2}(K_2)  
c_{\bar\alpha\sigma_1}(K_1),\non
\end{eqnarray}
where
$\ds
\prod_{i=1,2,3}\int_{-\infty}^{\infty}{d\e_{i}\over 2\pi}\int_{{\cal C}_{l}}{d\k_{i}\over (2\pi)^2}= 
$
$\ds
\int_{-\infty}^{\infty}{d\e_{1}\over 2\pi}
$
$\ds
\int_{-\infty}^{\infty}{d\e_{2}\over 2\pi}
$
$
\ds
\int_{-\infty}^{\infty}{d\e_{3}\over 2\pi}
$
$
\ds\int_{{\cal C}_{\bar\alpha;l}}{d\k_{1}\over (2\pi)^2}
$
$\ds
\int_{{\cal C}_{\alpha;l}}{d\k_{2}\over (2\pi)^2}
$
$\ds
\int_{{\cal C}_{\alpha;l}}{d\k_{3}\over (2\pi)^2}
$
$
\theta(\omega_{l}/2-\mid\!\!\varepsilon_{\bar\alpha}(k_{x4})\!\!\mid)
$
$
\theta(\Lambda-\mid \!\!k_{y4}\!\!\mid) 
$
with $\k_4=\k_1+\k_2-\k_3$.
The two-particle scattering processes considered here  
are depicted in Fig.~2.  Each leg of the scattering vertex is labeled by the regional index $\nu=\alpha,\bar\alpha$, 
the momentum in the $k_y$ direction and the spin.
The dimensionless  scattering  strengths  are given by
\begin{eqnarray}
g_l^{\sigma_1\sigma_2\sigma_3\sigma_4}(\kpara_1,\kpara_2,\kpara_3)=
\delta_{\sigma_4\sigma_2}\delta_{\sigma_3\sigma_1}g_l^{(1)}(\kpara_1,\kpara_2,\kpara_3)
-\delta_{\sigma_4\sigma_1}\delta_{\sigma_2\sigma_3}g_l^{(2)}(\kpara_1,\kpara_2,\kpara_3)
\end{eqnarray}
where  
$g_l^{(1)}(\kpara_1,\kpara_2,\kpara_3)$ and  $g_l^{(2)}(\kpara_1,\kpara_2,\kpara_3)$  denote the strengths of the
backward and forward 
scattering between the parallel flat regions, respectively. 
Within the present model, the electron umklapp process occurs in the special case where
 the distance between the parallel flat regions is equal to $\pi$.
In this work,  we do not take account of this case and neglect the electron umklapp process.

We split up the set of $\k$-points, ${\cal C}_{\nu;l}$, into  two subsets as ${\cal C}_{\nu;l}={\cal C}_{\nu;l+dl}^{<}\oplus d{\cal C}_{\nu;l+dl}^{>}$, where
$
{\cal C}_{\nu;l+dl}^{<}\equiv\left\{\k \mid \mid \varepsilon_\nu(k_x)\mid\leq \omega_{l+dl}/2, -\Lambda\leq k_y\leq \Lambda \right\}
$
(low-energy shell)
and
$
d{\cal C}_{\nu;l+dl}^{>}\equiv\left\{\k  \mid \omega_{l+dl} /2\leq \mid 
\varepsilon_\nu(k_x)\mid\leq \omega_{l}/2, -\Lambda\leq k_y\leq \Lambda\right\}
$
(high-energy shell).
Accordingly,   the     action is decomposed as
$
S_l=S^{<}_{l+dl}+S^{>}_{l+dl}.
$
Integration over the modes in the high-energy shell  gives the corresponding counterpart of the partition function as
\begin{eqnarray}
Z=\ds\int_{ {\cal C}_{l+dl}^{<}}{\cal D}c^{\ast} {\cal D}c \exp\left[S_{l+dl}^{<}+{\sum_{n=1}^{\infty}{{1\over n!}\langle\!\langle 
[S_{{\rm int};l+dl}^{>}]^{n}
\rangle\!\rangle_{\rm c}}}\right] 
,\label{eqn:avefast}
\end{eqnarray}
where
${\cal D}c^{\ast} {\cal D}c $ symbolizes the measure over the Fermion Grassmann variables
and ${\cal C}_{l+dl}^{<}$ means  that Fermion momenta   are restricted to the low-energy shell 
${\cal C}^<_{\alpha;l+dl}$ or ${\cal C}_{\bar\alpha;l+dl}^{<}$.
The average over the modes in the high-energy shell  is defined as
$
\langle\!\langle(\cdots)
\rangle\!\rangle\ds
=Z_{>}^{-1}\int_{ d{\cal C}_{l+dl}^>}{\cal D}c^{\ast} {\cal D} c \exp[S_{{\rm kin};l+dl}^{>}]\,\,
(\cdots), 
$
with 
$
Z_{>}=\ds\int_{ d{\cal C}_{l+dl}^>} {\cal D}c^{\ast} {\cal D} c  \exp[S_{{\rm kin};l+dl}^{>}]
$
and the subscript 'c' represents the connected diagrams.
We perform a perturbative expansion of 
${\sum_{n=1}^{\infty}{{1\over n!}\langle\!\langle 
[S_{{\rm int};l+dl}^{>}]^{n}
\rangle\!\rangle_{\rm c}}}$ 
by picking up the  Feynmann 
diagrams whose contribution is in proportion to $dl$ and then replacing
$S_{l+dl}^{<}+{\sum_{n=1}^{\infty}{{1\over n!}\langle\!\langle 
[S_{{\rm int};l+dl}^{>}]^{n}
\rangle\!\rangle_{\rm c}}}$  with
the renormalized action generally
written in a from
\begin{eqnarray}
&&\tilde S_{l+dl}^{<}=\ds\ds\sum_{\nu=\alpha,\bar\alpha}\sum_{\sigma}
\int_{-\infty}^{\infty}{d\e\over 2\pi}
\int_{{\cal C}_{\nu;l+dl}^<}{d\k\over (2\pi)^2}
[1+\theta_l(\kpara)dl+{\cal O}(dl^{2})]{\cal G}_{\nu}^{-1}(k_{x},\ii\e)  c_{\nu\sigma}^{\ast}(K)  c_{\nu\sigma}(K)\non\\
&&+{\pi v_{F}}
\sum_{\sigma_1,\dots,\sigma_4}\prod_{i=1,2,3}
\int_{-\infty}^{\infty}{d\e_i\over 2\pi}
\int_{{\cal C}_{l+dl}^{<}}{d\k_{i}\over (2\pi)^2} 
\left[g_l^{\sigma_1\sigma_2\sigma_3\sigma_4}(\kpara_1,\kpara_2,\kpara_3)\right.\label{eqn:REAC}\\
&&
\left.+w_l^{\sigma_1\sigma_2\sigma_3\sigma_4}(\kpara_1,\kpara_2,\kpara_3 ) dl
+{\cal O}(dl^{2})\right]
c_{\bar\alpha\sigma_4}^{\ast}(K_4)  
c_{\alpha\sigma_3}^{\ast}(K_3)  
c_{\alpha\sigma_2}(K_2)  
c_{\bar\alpha\sigma_1}(K_1).\non
\end{eqnarray}
The contribution 
$\theta_l(\kpara) {\cal G}_{\nu}^{-1}(k_{x},\ii\e) dl$ in the renormalized kinetic action 
comes from the two-loop self energy diagram  depicted in Fig.~3,
where $\theta_l(\kpara)$  is given by
\begin{eqnarray}
\theta_l(\kpara)={1\over 8}\left[
2g_l^{(1)}\Box
g_l^{(1)}
+2g_l^{(2)}\Box
g_l^{(2)}
-g_l^{(1)}\Box
g_l^{(2)}
-g_l^{(2)}\Box
g_l^{(1)}\right],\label{eqn:theta}
\end{eqnarray}
where the operation represented by $\Box$ is defined as
\begin{eqnarray}
g_l^{(i)} \Box g_l^{(j)} 
=\int_{-\Lambda+k_y}^{\Lambda+k_y}{dq_y\over 2\pi}\int_{-\Lambda-{\rm Min}(0,q_y)}^{\Lambda-{\rm Max}(0,q_y)}{dk_y'\over 2\pi}
g_l^{(i)}(k_{y}',k_{y},k_{y}-q_y )
g_l^{(j)}(k_{y}'+q_y,k_{y}-q_y,k_{y})\label{eqn:SE}.
\end{eqnarray}
The  integration ranges  are determined through  the
  condition that  all the incoming and outgoing momenta  at each scattering vertex in Fig.~3 must be
within the parallel flat regions. 
In the renormalized action for the two-particle scattering processes,
the contribution $w_{l}^{(i)}(\kpara_1,\kpara_2,\kpara_3 )dl$,
where 
$
w_l^{\sigma_1\sigma_2\sigma_3\sigma_4}(\kpara_1,\kpara_2,\kpara_3)=$
$
\delta_{\sigma_4\sigma_2}\delta_{\sigma_3\sigma_1}w_l^{(1)}(\kpara_1,\kpara_2,\kpara_3)
$
$
-\delta_{\sigma_4\sigma_1}\delta_{\sigma_2\sigma_3}w_l^{(2)}(\kpara_1,\kpara_2,\kpara_3),
$
 comes from the two-loop vertex correction diagrams  depicted in Fig.~4.
As space is limited, we do not write down full expression for     $w_{l}^{(i)}(\kpara_1,\kpara_2,\kpara_3 )$.

To restore the original cutoff,  we   rescale  the momenta and  frequencies as  
$
\tilde \k=(e^{dl}k_{x},k_y)=(\tilde k_x,k_y) 
$
and
$
\tilde \e=e^{dl}\e,
$
respectively.
Then,  to leave the kinetic action scale-invariant, we simultaneously perform the scale transformation of the Fermion Grassmann variable  as
$
\tilde   c_{\nu\sigma}(\tilde K)=[1+{1\over 2}\{\theta_l({\kpara}) -3\}dl]  c_{\nu\sigma}(K), 
$
where 
$\tilde K=( \tilde k_{x},k_{y},\ii \tilde\e)$.
Thus the two-loop RG equations for the scattering strengths are given by
\begin{eqnarray}
{d   g_l^{(i)}(\kpara_1,\kpara_2,\kpara_3 ) \over dl}=w_{l}^{(i)}(\kpara_1,\kpara_2,\kpara_3 )
-{1\over 2} g_l^{(i)}(\kpara_1,\kpara_2,\kpara_3 ) \sum_{j=1,\cdots,4}\theta_l(\kpara_j),
\label{eqn:VERTEX}
\end{eqnarray}
where $i=1,2$ and $k_{y4}=k_{y1}+k_{y2}-k_{y3}$.
The one-loop counterparts of the RG eqs.~(\ref{eqn:VERTEX}) are given by Zheleznyak, Yakovenko and Dzyaloshinskii,\cite{Zheleznyak97}
where $w_{l}^{(i)}(\kpara_1,\kpara_2,\kpara_3 )$ includes only one-loop vertex correction diagrams, Fig.~4(a) and 4(b),
  and the second term of the r.h.s of
the RG eq.~(\ref{eqn:VERTEX}) does not appear.
As an initial condition at $l=0$,  we put the scattering strengths    equal 
to the momentum-independent Hubbard repulsion, $U$,  i.e.
\begin{eqnarray}
g_0^{(i)}(\kpara_1,\kpara_2,\kpara_3 )=U/\pi v_F.\label{eqn:ICV}
\end{eqnarray}

The RG equation for the quasiparticle weight is obtained through  the renormalized kinetic action in eq.~(\ref{eqn:REAC})
and is written as
\begin{eqnarray}
{d \ln z_{l}(k_y)\over dl}= -\theta_l(\kpara). \label{eqn:QPW}
\end{eqnarray}
As an initial condition at   $l=0$,  we put the quasiparticle weight   equal 
to the non-interacting value, i.e.
\begin{eqnarray}
z_{0}(k_y)=1.\label{eqn:ICZ}
\end{eqnarray}
It is here instructive to see the $k_y$ dependence of  $z_{l}(k_y)$ by   assuming that
$g_l^{(i)}(\kpara_1,\kpara_2,\kpara_3 )$ remain at the initial value, $U/\pi v_F$. This  assumption applies to   the early-stage (i.e, high-energy)
 behavior of the RG flow.
Then,  eqs.~(\ref{eqn:theta}), (\ref{eqn:SE}) and  (\ref{eqn:QPW}) give the result,
$z_{l}(k_y)=[\omega_l/E_0]^{{\tilde U}^2(3\Lambda^2-k_y^2)/16\pi^2 }$ with $\tilde U=U/\pi v_F$, which
 indicates that the quasiparticle weight at the center of the flat region ($k_y=0$) is the most strongly suppressed due solely to
{\it the kinematical restriction to the logarithmically divergent processes}.
In reality,  however,  we cannot obtain the  2-loop RG flow of $z_{l}(k_y)$ in a consistent manner, until we  take account of
 {\it the anisotropic   RG flow of the scattering strengths.}

We   solve numerically the  RG equations,
(\ref{eqn:VERTEX}) and (\ref{eqn:QPW}), under the initial conditions, (\ref{eqn:ICV}) and (\ref{eqn:ICZ}).
Absolute value of the Hubbard repulsion, $U$, is not essential for the RG flow of the scattering strengths and the quasiparticle weight.
From now on, we set the Hubbard repulsion at $U=4.5\pi v_F$.  
Linear differential equations, (\ref{eqn:VERTEX}) and (\ref{eqn:QPW}),  are treated as recursion equations  with an infinitesimal difference step $dl=0.08$. 
The numerical procedure at the $n$-th step with the corresponding scaling parameter $l_n=ndl$   consists of three consecutive steps.
First,
 we  insert $g_{l_n}^{(i)}(\kpara_1,\kpara_2,\kpara_3 )$ into the expressions of 
$\theta_{l_n}(\kpara)$ and $w_{l_n}^{(i)}(\kpara_1,\kpara_2,\kpara_3 )$.
Second,  we perform the integration
over the momenta in the $k_y$ direction by  dividing the interval, $-\Lambda\leq k_y\leq \Lambda$, into 32 discrete points
to obtain $\theta_{l_n}(\kpara)$ and $w_{l_n}^{(i)}(\kpara_1,\kpara_2,\kpara_3 )$.
Finally, we obtain $g_{l_{n+1}}^{(i)}(\kpara_1,\kpara_2,\kpara_3 )$   by using eq.~(\ref{eqn:VERTEX}). 

In Fig.~5(a), we show the RG flow of the quasiparticle weight, $z_{l}(k_y)$,   for $0\leq l\leq 3.2$. 
When the energy scale decreases, the  quasiparticle weight is the most strongly suppressed   around the center of the flat region, $k_y=0$,
forming a dip there.
In Fig.~5(b), we show the same RG flow   for $3.2\leq l\leq 4.08$. The $k_y$ dependence of the quasiparticle
weight undergoes a crossover behavior toward its low-energy asymptotics.  There is some crossover scaling parameter $l_{\rm cross}\sim 3$ around which
the   dip   of $z_{l}(k_y)$ around $k_y=0$ becomes flatter. Then,  in the   low-energy regime,  $l>l_{\rm cross}$,  
the region  around which $z_{l}(k_y)$ is the most strongly suppressed moves  from the center  ($k_y=0$) toward the edges ($k_y=\pm \Lambda$) and 
finally $z_{l}(k_y)$ approaches zero everywhere for $-\Lambda\leq k_y \leq\Lambda$ in the low-energy limit.
The $k_y$-dependence of the quasiparticle weight in the high-energy regime corresponds to the anisotropic pseudogap behavior in
underdoped Bi$_2$Sr$_2$CaCu$_2$O$_8$ suggested by the ARPES experiments.\cite{Norman98,Ding96}  
In the present analysis, the crossover value of the scaling parameter, $l_{\rm cross}$, corresponds to the temperature scale, $T_{\rm cross}\sim 0.05 E_0$.
Thus we may expect that the temperature range coverved by the ARPES experiments on the normal state of Bi$_2$Sr$_2$CaCu$_2$O$_8$ lies 
within the high-energy regime introduced here.
We here again stress that, in the present scheme, the anisotropic RG flow of the quasipartcile weight occurs due solely to the logarithmically
singular processes coming from the flat regions of the Fermi surface.

To see the reason why the crossover behavior of $z_{l}(k_y)$ occurs, in Fig.~6, we show the RG flow of the backward scattering strength,
$g_{l}^{(1)}(\kpara_1,\kpara_2,\kpara_3 )$ at     
points, $(\kpara_1,\kpara_2,\kpara_3 )$ $=(0,0,0)$,  $(\Lambda,-\Lambda,\Lambda)$,
$(\Lambda,-\Lambda,-\Lambda)$, $(\Lambda,\Lambda,\Lambda)$.  
As an example, let us see   the RG flow of    $g_{l}^{(1)}(0,0,0)$.  The monotonic decrease of $g_{l}^{(1)}(0,0,0)$  at the early stage  of the renormalization
reminds us of the RG flow of the backward scattering strength in the purely one-dimensional metal, where
the   backward scattering strength monotonously decreases and vanishes toward the low-energy asymptotics corresponding to
the Tomonaga-Luttinger fixed point.\cite{Solyom79} 
In the present case in contrast to this,  the RG flow of $g_{l}^{(1)}(0,0,0)$  has a very shallow minimum around $l\sim l^\ast$
 and begins to increase for $l>l_{\rm cross}$.  
As is seen in  the inset in Fig.~6,    the flow becomes nearly constant for $l>l_{\rm c}\sim 4.1$  in the   low-energy limit 
due to the two-loop vertex corrections, while, in the case of the one-loop RG analysis,  the corresponding 
flow  diverges around $l\sim l_{c}$.\cite{Zheleznyak97}

The crossover behavior of $g_{l}^{(1)}(0,0,0)$   originates
from   breakdown of the cancellation between the particle-particle (Cooper) loop[Fig.~4(a)] and the particle-hole (Peierls) loop[Fig.~4(b)]. 
As the renormalization process goes on, the breakdown becomes more and
more conspicuous and the Peierls loop tends to dominate the Cooper loop. 
Consequently,  the spin-density-wave fluctuation becomes the most dominant fluctuation in the low-energy limit, provided that  
the  regions $\alpha$ and  $\bar \alpha$ are perfectly flat.\cite{Zheleznyak97}
The crossover behavior of the RG flow of $z_{l}(k_y)$ corresponds to the crossover behavior of the scattering strengths from the high-energy
  regime  to the low-energy   regime.

We here consider the effects of   small but finite curvature of the Fermi surface in the regions $\alpha$ and  $\bar \alpha$ which always exists in reality.
  In this case, 
at the energy scale smaller  than  the energy resolution, $\Delta E$,  which detects  the Fermi surface curvature,
  the two-loop self-energy diagram never  gives rise to
the contribution of the type
$\theta_l(\kpara) {\cal G}_{\nu}^{-1}(k_{x},\ii\e) dl$ in the renormalized kinetic action in eq.(\ref{eqn:REAC}).
Consequently,
{\it the RG flow of $z_{l}(k_y)$ stops} around the scaling parameter, $l_{\rm curv}$, defined
 as $\Delta E\sim E_0 e^{-l_{\rm curv}}$.
In Fig.~7, 
we take $l_{\rm curv}=3$ (corresponding to $\Delta E/E_0\sim 0.05$) as an example and show the RG flows of $z_{l}(k_y)$ at some representative points on the Fermi surface.
 As the energy scale (or temperature scale) decreases, the quasiparticle weight at the point {\bf a} on the round-arc region 
remains marginal, indicating that 
 the round-arc regions of the Fermi surface  are robust.
 In the  flat regions,  the quasiparticle weight at the center, {\bf c}, is more strongly suppressed than that at the edge,  {\bf b},
 but remains finite in the low-energy limit due to the curvature of the Fermi surface.

In our analysis, the RG flow of   $z_{l}(k_y)$ becomes discontinuous at the edge points  of the flat region (the point {\bf b} in the inset of the Fig.~7), 
which seems unnatural.
We give here a qualitative comment on the origin of this discontinuity.  
In the present   analysis, we treated the flat regions as decoupled to the round-arc regions
in the RG sense.
However, strictly speaking,  at  finite energy scale,  
the flat regions and the round-arc regions can couple to each other through  the scattering processes involving the very tiny region,  the shaded area in Fig~8,
which can be regarded as  an extension of the flat region.   
Thus, if we take account of these processes, the discontinuity of $z_{l}(k_y)$
at the edge points  will be removed.  
 We do not  inquire further into this point, since this superficial discontinuity is irrelevant
to our qualitative understanding of the  RG flow of $z_{l}(k_y)$ in the present model.  

In summary,  we have carried out   the two-loop renormalization-group analysis for
the two-dimensional electron system with the partially flat Fermi surface and  found that
the logarithmically singular processes coming from the flat regions cause the  renormalization-group flow
 of the quasiparticle weight which depends on the location of the $\k$-point in the flat regions of the Fermi surface.
When the energy scale decreases,   the quasiparticle weight is the most strongly suppressed   around the center of the flat region, forming a dip there
which
 qualitatively agrees with the
anisotropic pseudogap behavior suggested through the  ARPES  experiments.

J.K. was   supported by a Grant-in-Aid for Encouragement of Young Scientists   from the Ministry of Education, Science, Sports and Culture, Japan.

\begin{figure}
\caption{The model Fermi surface considered here.}
\end{figure}

\begin{figure}
\caption{The two-particle scattering vertex considered here.
The 
solid and broken lines represent the propagators for the  electrons in the flat regions, $\alpha$ and $\bar \alpha$, respectively.
The box represents the   scattering vertex,  $g_l^{\sigma_1\sigma_2\sigma_3\sigma_4}(\kpara_1,\kpara_2,\kpara_3)$.}
\end{figure}

\begin{figure}
\caption{The two-loop self-energy diagram.}
\end{figure}

\begin{figure}
\caption{The diagrams which contribute,
(a) to   the one-loop vertex correction  in the particle-particle (Cooper) channel,   
(b) to   the one-loop vertex correction  in the particle-hole (Peierls) channel,
and 
(c) to   the two-loop vertex corrections.}
\end{figure}

\begin{figure}
\caption{$k_y$ dependence of the renormalization-group flow of the quasiparticle weight, $z_{l}(k_y)$,  
(a) for $0\leq l\leq 3.2$, and
(b)   for $3.2\leq l\leq 4.08$. }
\end{figure}

\begin{figure}
\caption{The renormalization-group flow of the backward scattering strength,
$g_{l}^{(1)}(\kpara_1,\kpara_2,\kpara_3 )$ at  $(\kpara_1,\kpara_2,\kpara_3 )$ $=(0,0,0)$,  $(\Lambda,-\Lambda,\Lambda)$,
$(\Lambda,-\Lambda,-\Lambda)$, $(\Lambda,\Lambda,\Lambda)$. 
In the inset,    the same graph covering   a wider range of the vertical axis is shown.}
\end{figure}

\begin{figure}
\caption{The renormalization-group flow of the quasiparticle weight, $z_{l}(k_y)$, at some representative points on the Fermi surface.}
\label{reduced}
\end{figure}

\begin{figure}
\caption{The shaded area also contributes to the renormalization-group flow of the quasiparticle weight, which is not taken into account
in this study.
 }
\end{figure}

\begin{references}
\bibitem{Imada98}See, for example, M. Imada, A. Fujimori, and Y. Tokura,  to appear in Rev. Mod. Phys. (1998).
\bibitem{Norman98}M. R. Norman, H. Ding, M. Randeria, J. C. Campuzano, T. Yokoya, T. Takeuchi, T. Takahashi, T. Mochiku, K. Kadowaki, P. Guptasarma, and D. G. Hinks,  Nature\  {\bf 392}, 157 (1998).
\bibitem{Ding96}H. Ding, T. Yokoya, J. C. Campuzano, T. Takahashi, M. Randeria, M. R. Norman, T. Mochiku,
K. Kadowaki, and J. Giapintzakis,  Nature\  {\bf 382}, 51 (1996); 
H. Ding, M. R. Norman, T. Yokoya, T. Takeuchi, M. Randeria, J. C. Campuzano, T. Takahashi, T. Mochiky, and K. Kadowaki, {\prl}\  {\bf 78}, 2628 (1997).
\bibitem{Engelbrecht97}J. R. Engelbrecht, A. Nazarenko,  M. Randeria, and E. Dagotto, {\prb}\   {\bf 57}, 13406 (1998).
\bibitem{Wen96}X. G. Wen and P. A. Lee, {\prl}\   {\bf 76}, 503 (1996).
\bibitem{Shankar94} R. Shankar, Rev. Mod. Phys. \ {\bf 66}, 129 (1994).
\bibitem{Solyom79}J. S\'olyom, Adv. Phys. {\bf 28}, 201 (1979).
\bibitem{Marshall96}D. S. Marshall {\it et al.},  {\prl}\  {\bf 76}, 4841 (1996); A. G. Loeser {\it et al.},  Science\  {\bf 273}, 325 (1996).
\bibitem{Zheleznyak97}A. T. Zheleznyak V. M. Yakovenko, and I. E. Dzyaloshinskii, {\prb}\   {\bf 55}, 3200 (1997).
\end{references}
\end{document}